# Buffer-layer-controlled Nickeline vs Zinc-Blende/Wurtzite-type MnTe growths on c-plane $Al_2O_3$ substrates

Deepti Jain[1], Hee Taek Yi[1], Alessandro R. Mazza[2,3], Kim Kisslinger[4], Myung-Geun Han[5], Matthew Brahlek[2] and Seongshik Oh[1,6,*]

[1] Department of Physics and Astronomy, Rutgers, The State University of New Jersey, Piscataway, NJ 08854, USA

[2] Materials Science and Technology Division, Oak Ridge National Laboratory, Oak Ridge, TN 37831, USA

[3] Present address: Center for Integrated Nanotechnologies, Los Alamos National Laboratory, Los Alamos, New Mexico 87545, USA

[4] Center for Functional Nanomaterials, Brookhaven National Laboratory, Upton, NY 11973, USA

[5] Condensed Matter Physics and Materials Science, Brookhaven National Laboratory, Upton, NY 11973, USA

[6] Center for Quantum Materials Synthesis, Rutgers, The State University of New Jersey, Piscataway, NJ 08854, USA

[*] Correspondence should be addressed to ohsean@physics.rutgers.edu

**Abstract**

In the recent past, MnTe has proven to be a crucial component of the intrinsic magnetic topological insulator (IMTI) family $[MnTe]_m[Bi_2Te_3]_n$, which hosts a wide range of magneto-topological properties depending on the choice of *m* and *n*. However, bulk crystal growth allows only a few combinations of *m* and *n* for these IMTIs due to the strict limitations of the thermodynamic growth conditions. One way to overcome this challenge is to utilize atomic layer-by-layer molecular beam epitaxy (MBE) technique, which allows arbitrary sequences of $[MnTe]_m$ and $[Bi_2Te_3]_n$ to be formed beyond the thermodynamic limit. For such MBE growth, finding optimal growth templates and conditions for the parent building block, MnTe, is a key requirement. Here, we report that two different hexagonal phases of MnTe - nickeline (NC) and zinc-blende/wurtzite (ZB-WZ) structures, with distinct in-plane lattice constants of 4.20 ± 0.04 Å and 4.39 ± 0.04 Å, respectively – can be selectively grown on c-plane $Al_2O_3$ substrates using different buffer layers and growth temperatures. Moreover, we provide the first comparative studies of different MnTe phases using atomic-resolution scanning transmission electron microscopy and show that ZB and WZ-like stacking sequences can easily alternate between the two. Surprisingly, $In_2Se_3$ buffer layer, despite its lattice constant (4.02 Å) being closer to that of the NC phase, fosters the ZB-WZ instead, whereas $Bi_2Te_3$, sharing the same lattice constant (4.39 Å) with the ZB-WZ phase, fosters the NC phase. These discoveries suggest that lattice matching is not always the most critical factor determining the preferred phase during epitaxial growth. Overall, this will deepen our understanding of epitaxial growth modes for chalcogenide materials and accelerate progress toward new IMTI phases as well as other magneto-topological applications.

## I. Introduction

MnTe has been studied extensively in the past few decades due to its innate property of being an antiferromagnetic semiconductor, making it valuable for memory devices [1,2], optoelectronics [3,4] and spintronics [5-7] to name a few. It is known to exist in three phases; nickeline (NC, Fig. 1(a,b)), wurtzite (WZ, Fig. 1(c,d)), and zinc blende (ZB, Fig. 1(e,f)) [8]. In bulk crystals, which are governed by thermodynamics, MnTe crystallizes only in hexagonal NC structure. Consequently, NC-structure MnTe phase has been most comprehensively investigated, and its structural, optical, electronic and magnetic properties are well established [7,9-16]. Due to its high Neel temperature of 307-310K [9,10], it is a candidate for antiferromagnetic spintronics applications that can be operated at room temperature. It is worth noting that while NC MnTe has always been identified as an antiferromagnet, it was recently predicted to belong to a new and distinct class of magnetic materials known as altermagnets [17,18]. The other phases of MnTe, however, are metastable; they cannot be grown via equilibrium growth methods and require different modes of growth to materialize.

Interest in the ZB MnTe phase stemmed from its role in some diluted magnetic semiconductors and it was found that it has a lower Neel temperature (~65 K) [19,20] and a wider optical bandgap (~3 eV) [21-24] compared to the NC phase (1.26-1.5 eV) [14,15]. While majority of ZB MnTe has been grown via MBE, with Mn being deposited under excess Te, there have also been a few reports in which the ionized cluster beam method was used, with NC MnTe polycrystals as the source material. Cubic ZB MnTe has mostly been achieved by epitaxial stabilization on ZB-structure substrates such as GaAs, CdTe and InSb, often utilizing additional buffer layers like ZnTe and CdTe between the substrate and the film [19,21,25-27]. When ZB-structure is viewed along the (111) direction, the arrangement of atoms on the surface has six-fold symmetry, as can be seen from Fig. 1(e), and the hexagonal-like ZB MnTe(111) film has also been stabilized on GaAs(001) [25,26], $BaF_2$(111) [24], $SrTiO_3$(001) [28], mica [29] and $Al_2O_3$(0001) with CdTe(111) buffer [20]. On the other hand, intrinsically hexagonal WZ MnTe phase [Fig. 1(c)], with a wide bandgap of 2.4-3 eV, has been so far grown only in a polycrystalline form on amorphous substrates like glass and indium-zinc-oxide, for optoelectronic applications [2,4,30,31]. Sometimes, multi-phases of MnTe have been observed depending on the choice of substrate temperature and Te:Mn flux ratio, and they usually involve the stable phase, NC MnTe, and either of the metastable phases, ZB or WZ MnTe [4,28-30]. However, co-existing ZB and WZ MnTe phases have never been reported, despite their structural similarities.

One of the active research areas where MnTe plays a critical role is the newly discovered intrinsic magnetic topological insulator (IMTI) family, $[MnTe]_m[Bi_2Te_3]_n$ [32-35], a class of materials that has been predicted to host exotic topological phases such as axion insulators [36,37], magnetic Weyl semimetals [33,38], and high temperature quantum anomalous Hall effects [32,39]. Despite some progresses, there are still multiple challenges hindering the realization of its many possibilities. So far, majority of studies on $[MnTe]_m[Bi_2Te_3]_n$ have been performed on bulk crystals and thin flakes exfoliated from them. Inherently, bulk crystal growth relies on macroscopic diffusion of constituent elements along all the three directions until they reach thermodynamically the most stable configuration. In such a growth mode, it becomes extremely difficult to form highly layered structures with large unit cell sizes, as in the case of [MnTe]$_m$[Bi$_2$Te$_3$]$_n$ compounds with large $m$ and $n$ values. Additionally, among the compounds that are thermodynamically stable i.e., $m = 1$, $n \geq 1$, the ones with higher $n$ exist within a very narrow range of temperatures. As a consequence, so far only $m = 1$, $n \leq 7$ phases have successfully been grown [34]. These shortcomings can be overcome by growing the films with atomic-layer-by-layer molecular beam epitaxy (MBE) technique, which can potentially enable finely tuned, atomic scale engineering of this family of materials for all possible values of $m$ and $n$. The main motivation behind the following work is creating a foundation to grow thin films of these IMTIs, by focusing on finding a suitable template to grow one of its building blocks first, i.e., MnTe. Here, we report that on Al$_2$O$_3$(0001) substrates, which are both economical and successfully used for various high quality topological thin film growths [40-44], NC and ZB-WZ phases of MnTe can be selectively grown with relative ease, using two distinct buffer layers combined with different growth temperatures.

## II. Methods

On Al$_2$O$_3$(0001) substrates, our initial goal was to find optimal growth conditions for NC MnTe phase because it is the phase found in the bulk crystals, and the atomic-sequence of Te atoms surrounding the Mn layer in bulk crystals of [MnTe][Bi$_2$Te$_3$]$_n$ is equivalent to that of the NC structure [45]. Prior to any deposition, the substrates were cleaned ex-situ by UV generated ozone followed by in-situ heating up to 750 °C under oxygen pressure of $1 \times 10^{-6}$ Torr. This step helps get rid of any organic contaminants on the surface of the substrate. Thickness of the films were determined by quartz crystal microbalance (QCM) and Rutherford backscattering spectroscopy (RBS), and the growth was monitored in-situ using reflection high-energy electron diffraction (RHEED). For all the films described below, adsorption-controlled growth mode is used with several times more tellurium or selenium fluxes than those of the metal elements. Each element is evaporated from a standard effusion cell.

The lattice constant of NC MnTe, $a_{NC} = 4.14$ Å, is not a good match with that of the sapphire substrate, $a_{Al2O3} = 4.76$ Å. A common solution for such a lattice mismatch problem is to introduce a structurally compatible buffer layer between the film and the substrate. In this case, the buffer layer chosen was insulating In$_2$Se$_3$, with a lattice constant of a$_{In2Se3}$ = 4.02 Å, which is closer to that of NC MnTe. We have previously grown high-quality, single-phase In$_2$Se$_3$ on Al$_2$O$_3$, involving a multi-step recipe [42]. Accordingly, first, a seed layer of 3 QL Bi$_2$Se$_3$ is grown at 135 °C and an additional 7 QL is deposited at 300 °C. This serves as a good template to grow In$_2$Se$_3$ (5 nm), also at 300 °C. When this layered structure is annealed to 600 °C, the Bi$_2$Se$_3$ layer diffuses through the In$_2$Se$_3$ and evaporates away, leaving behind In$_2$Se$_3$ directly on Al$_2$O$_3$. Once the buffer layer is ready, the substrate is cooled down to 450 °C and the MnTe film is grown on top [Fig. 2(a)]. The final growth temperature was chosen after multiple trials, based upon the temperature range in which a single crystalline, 2D surface could be observed with RHEED.

Contrary to our expectation, the phase of MnTe grown on In$_2$Se$_3$ was not NC. Fig. 3(a) shows snapshots of RHEED patterns of the In$_2$Se$_3$ buffer layer and the MnTe film taken during growth: the sharp and localized streaks imply a good epitaxial growth for both, the buffer and the film. Based on the RHEED streak spacing, the in-plane lattice constant of the MnTe film is found to be 4.39 ± 0.04 Å, quite different from the expected 4.14 Å of NC MnTe. Additionally, on comparing the distance between RHEED streaks for the two high symmetry directions, it can be seen that the distance ratio is $\sqrt{3}$, which is an indication that it has six-fold in-plane symmetry. According to the literature, it could be either ZB(111) or WZ(0001) phase.

Since we were unable to obtain NC MnTe on In$_2$Se$_3$, we tried using a different buffer layer: Bi$_2$Te$_3$. As can be inferred from the existence of [MnTe]$_m$[Bi$_2$Te$_3$]$_n$, Bi$_2$Te$_3$ and NC MnTe are compatible with each other. However, Bi$_2$Te$_3$ is conducting; hence, we would need a very thin layer of it, so that it does not interfere with the transport properties of MnTe. Based on our previous reports, Bi$_2$Te$_3$ has poor adhesion to inert Al$_2$O$_3$ substrate at its optimal growth temperature, and inserting a less inert layer like Cr$_2$O$_3$ between them can help it stick better [44]. Following this recipe, 1 nm Cr$_2$O$_3$ was deposited on the substrate at 700 °C under oxygen pressure of $1 \times 10^{-6}$ Torr and after that 1 QL Bi$_2$Te$_3$ was grown at 300 °C. We finally grew MnTe on this template at 300 °C [Fig. 2(b)]. We can see from Fig. 3(b) that the in-plane lattice constant of MnTe grown on this buffer is 4.20 ± 0.04 Å, much closer to that of NC MnTe. In Fig. 3(c) and (d), we can see that while both types of MnTe phases start growing with a lattice constant close to that of their respective buffer layer, they gradually relax to the final lattice constants of 4.39 Å and 4.20 Å, respectively, after 10~20 monolayers.

## III. Results and Discussion

To probe further, x-ray diffraction (XRD) was carried out on the two films, using a Panalytical X'Pert Pro and a monochromated Cu $K_{\alpha 1}$ source. Fig. 4 shows the 2θ scans for both MnTe films. Signature (0003n) peaks belonging to the $Al_2O_3$ substrate and a small peak from the Se capping layer can be seen in the patterns of both films. All the peaks of the MnTe film with lattice constant 4.20 ± 0.04 Å can be identified with the (0002n) peaks of NC MnTe phase in the literature, as shown in Fig. 4(c) [46]: this implies that this is in fact the NC MnTe phase. On the other hand, the XRD pattern of the MnTe film grown on $In_2Se_3$ buffer reveals the coexistence of both ZB and WZ phases. It can be seen in Fig. 4(a) that each prominent peak is a superposition of two peaks; ZB(nnn) and WZ(0002n). Fig. 4(b) shows an enlarged section of the XRD pattern where the ZB(111) and WZ(0002) peaks have been resolved, and are consistent with similar studies conducted previously [28,31]. The presence of these two peaks also implies that the c-axis lattice constant is not uniform for the entire film. Transport studies involving temperature dependence of longitudinal resistance of a 30 nm NC MnTe film are shown in Fig. 4(d). The shape indicates predominantly semiconducting behavior, similar to previous reports on NC MnTe. A small, elongated hump can be seen around 250-300 K consistent with an AFM transition temperature associated with interaction between itinerant electrons and localized Mn spins [47]. While we could not perform similar measurements on the ZB-WZ MnTe phase, we carried out I-V measurements and found the two-point resistance of the ZB-WZ phase to be ~20 GΩ at room temperature. This is not surprising since the reported bandgaps for ZB MnTe and WZ MnTe are ~3 eV and 2.4-3 eV respectively, compared to 1.26-1.5 eV for NC MnTe.

High-angle annular dark-field scanning tunneling electron microscopy (HAADF-STEM) was carried out to gain insight into the atomic structure of these films. The cross-sectional STEM sample was prepared using a FEI Helios G5 UX focused ion beam system with final Ga+ milling performed at 2 keV. Then, the HAADF-STEM was performed with a JEOL ARM 200CF equipped with a cold field emission gun and spherical aberration correctors, which was operated at 200 kV. The detection angles for HAADF imaging were ranging from 68 to 280 mrad. Figures 5(a) and (e) show clear boundaries between the films and the buffer layers. Upon closer inspection, it can be seen that there is a difference in arrangement of Mn and Te atoms in both films. In Fig. 5(f), the Mn atoms fall in a straight line along the (0001) direction while the Te atoms form a zigzag pattern. This AcBcAcBc sequence is characteristic of the NC structure due to interpenetrating primitive hexagonal lattice of the Mn atoms and close packed hexagonal lattice of Te atoms [Fig. 1(b)]. In contrast, the Mn atoms do not align along the direction of growth in ZB-WZ MnTe [Fig. 5(b)]. Additionally, on comparing the positions of Mn and Te atoms relative to one another in the STEM image of the ZB-WZ phase [Fig. 5(b)], Fig. 1(d) and Fig. 1(f), it can be seen that the directions of growth are $(000\overline{1})$ and $(\overline{111})$ for the WZ and ZB phase respectively. Fig. 5(c) and (d) further illustrate the arrangement of atoms in ZB-WZ MnTe. It can be seen that the initial growth of ZB-WZ MnTe follows an ABCABC sequence, corresponding to the $(\overline{111})$ growth mode of ZB structure illustrated in Fig. 1(f). However, as highlighted by the yellow dotted lines in Fig 5(c), stacking faults are very common and the sequence sometimes changes to ABAB, which corresponds to the WZ stacking shown in Fig. 1(d). As the growth progresses, the stacking changes between ZB and WZ quite randomly [Fig. 5(d)]. It can be inferred from the STEM images that ZB MnTe(111) and WZ MnTe have similar formation energies. Nonetheless, despite frequent switching between ZB and WZ stacking along the direction of growth, it is notable that the in-plane lattice constant as judged from the RHEED approaches the stable value of 4.39 ± 0.04 Å, which is identical to that of $Bi_2Te_3$ as shown in Fig. 3(a,c). Here, it is important to note that this ZB-WZ-sequence mixed phase is very different from other common mixed phases, in that the only difference between ZB and WZ phases is just the stacking sequence. In other words, from the viewpoint of the topmost layer, this mixed-sequence structure is just like a single-crystalline 2D lattice structure with a well-defined lattice constant of 4.39 ± 0.04 Å, slightly smaller than the effective lattice constants (~4.5 Å) [21,48] of pure ZB or WZ phases. Accordingly, this ZB-WZ mixed-sequence platform could provide its own unique applications distinct from the pure ZB or WZ platform.

## IV. Conclusion

Our study provides a detailed structural analysis of NC and ZB-WZ phases of MnTe, grown selectively on $Al_2O_3$(0001) substrates with two different buffer layers. Surprisingly, $In_2Se_3$ buffer layer, despite its lattice constant (4.02 Å) being closer to that of the NC phase, fosters the ZB/WZ phase, whereas $Bi_2Te_3$, sharing the same lattice constant (4.39 Å) with the ZB/WZ phase, fosters the NC phase. This suggests that lattice matching is not always the most critical factor determining the preferred phase during epitaxial growth. Furthermore, we have provided the first atomic-resolution STEM studies of ZB-WZ MnTe phase, showing that stacking sequences of ZB(111) and WZ(0001) MnTe can easily alternate from each other, which suggests that the formation energies of ZB(111) and WZ(0001) MnTe are extremely close to each other. The absence of literature for the coexistence of these two phases, despite their similar formation energies, could be due to two possibilities. First, the few reports of ZB(111) MnTe growth in the literature may have some portions of WZ-like stacking, and vice versa, but could not be confirmed due to lack of detailed STEM studies. Second, the $In_2Se_3$ buffer layer may play a role in stabilizing both of these meta-stable phases. Although it is an open question whether it is possible to achieve pure ZB or WZ sequences on the $In_2Se_3$ buffer layer, having MnTe thin films with ZB-WZ mixed sequences on the $In_2Se_3$ buffer is an unexpected yet significant finding, considering the proximity of its in-plane lattice constant (4.39 ± 0.04 Å) to that (4.39 Å) of $Bi_2Te_3$ and $[MnTe]_m[Bi_2Te_3]_n$. Due to this perfect lattice match, it can be used as an optimal foundation to grow these IMTIs with minimal interfacial defects. As demonstrated recently by some of us, this also opens the possibility of combining superconducting Fe(Te,Se) with IMTIs toward novel topological superconductivity, via hybrid symmetry epitaxy [49]. On the other hand, the NC phase with $Bi_2Te_3$-$Cr_2O_3$ buffer layer can also be tailored to enable growth of $[MnTe]_m[Bi_2Te_3]_n$ on

$Al_2O_3$(0001) with different interfacial conditions, as will soon be published in a follow-up work. These studies shed light on the critical role of buffer layers in stabilizing selected phases and will open many avenues in topological and spintronic applications.

**Acknowledgements**

This work is supported by National Science Foundation's DMR2004125, Army Research Office's W911NF2010108, and MURI W911NF2020166. The work at Oak Ridge National Laboratory is supported by the U. S. Department of Energy (DOE), Office of Science, Basic Energy Sciences (BES), Materials Sciences and Engineering Division. The scanning transmission electron microscopy work performed at Brookhaven National Laboratory is sponsored by the US Department of Energy, Basic Energy Sciences, Materials Sciences and Engineering Division, under contract no. DE-SC0012704. This research used the Electron Microscopy resources (the Helios G5 FIB) of the Center for Functional Nanomaterials (CFN), which is a U.S. Department of Energy Office of Science User Facility, at Brookhaven National Laboratory under Contract No. DE-SC0012704. We acknowledge Hussein Hijazi for RBS measurements.

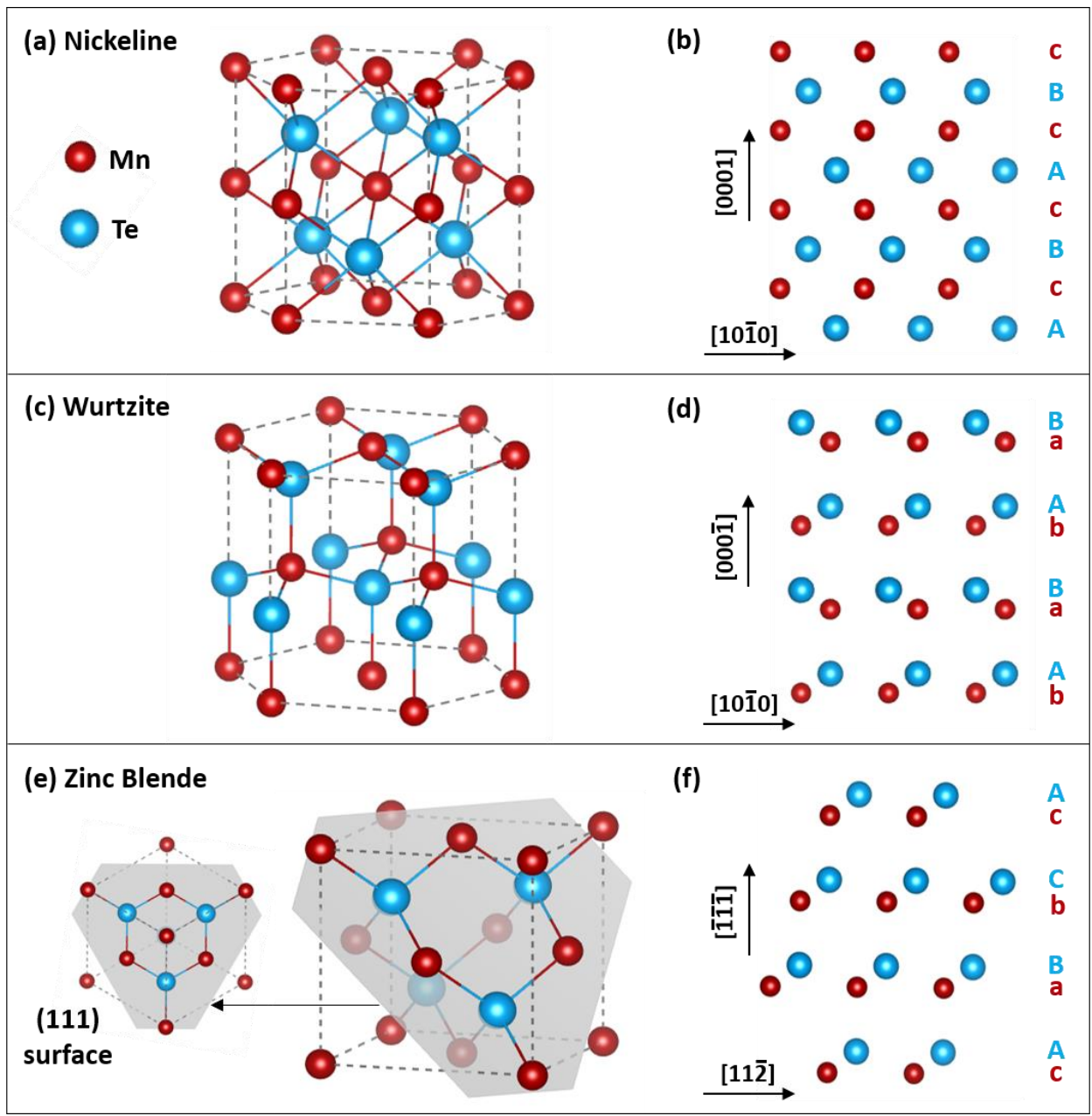

**Figure 1. Crystal structures of different phases of MnTe. (a,c,e)** Nickeline, wurtzite and zinc blende phases of MnTe respectively. The hexagonal-like (111) plane which is the surface in (111) growth mode of the zinc blende phase is illustrated in (c). **(b,d,f)** Differences in stacking sequences for Mn and Te atoms in the nickeline, wurtzite and zinc blende phase respectively: the crystallographic directions are chosen to match those in Fig. 5.

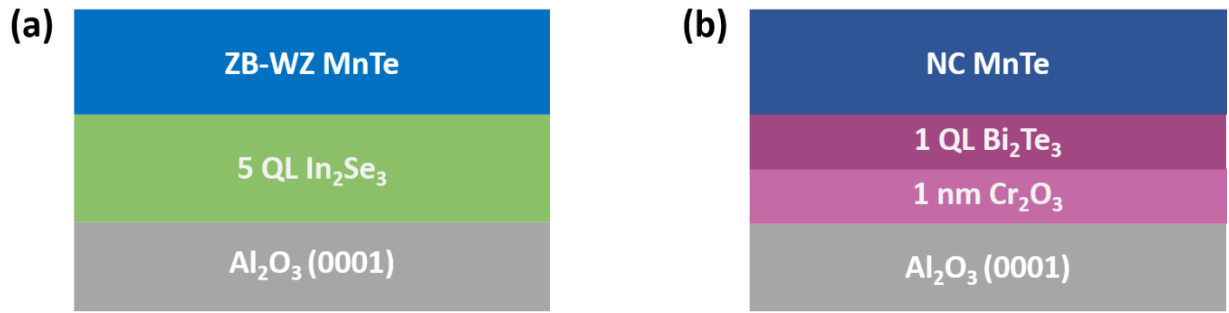

**Figure 2. Growth schematic.** A schematic illustrating the buffer layered growth of **(a)** ZB-WZ MnTe and **(b)** NC MnTe on Al$_2$O$_3$(0001) substrates.

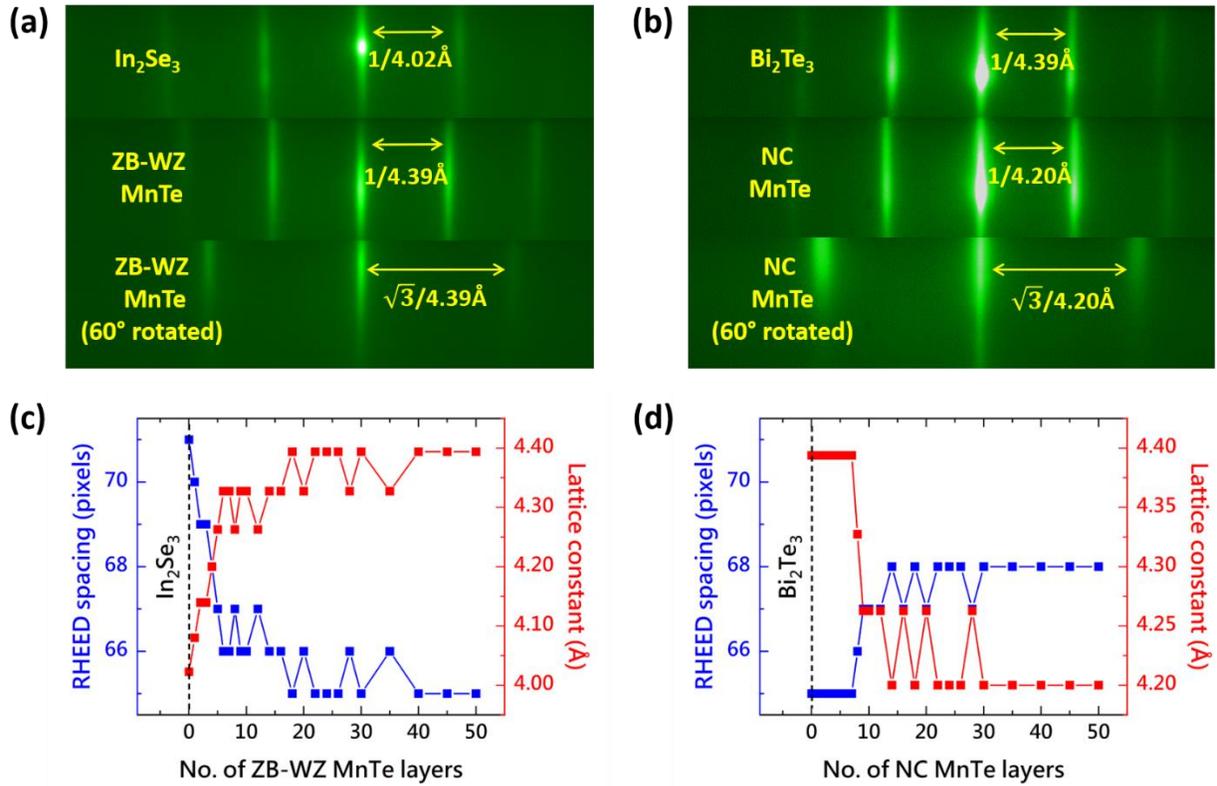

**Figure 3. RHEED analysis of the two different phases. (a,b)** RHEED patterns showing real time growth of epitaxial (a) ZB-WZ MnTe and (b) NC MnTe. Both lattice constants are highlighted with respect to that of their buffer layers. Additionally, by comparing the streaks for the high symmetry directions, it can be seen that the in-plane structures for both films are hexagonal, as indicated by the geometric ratios. **(c,d)** Evolution of growth for (c) ZB-WZ MnTe and (d) NC MnTe as seen via RHEED spacing in pixels, inversely proportional to the lattice constant. Due to the discrete nature of pixels, the error bar for the lattice constant is ~0.04 Å.

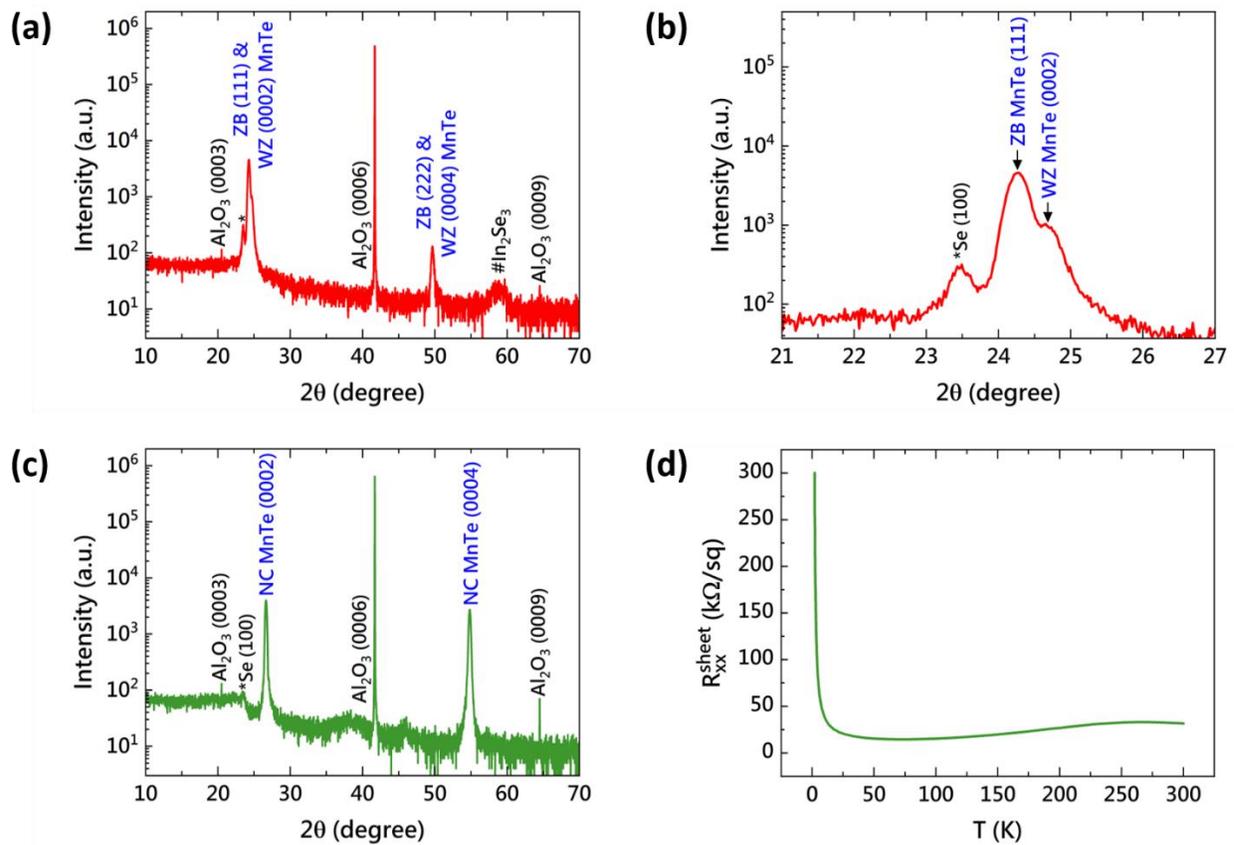

**Figure 4. X-Ray Diffraction patterns and R vs T.** (**a**) XRD pattern of ZB-WZ MnTe with ZB(nnn) and WZ(0002n) peaks highlighted. (**b**) Magnified portion of (a) showing ZB(111) and WZ(0002) peaks. (**c**) XRD pattern of NC MnTe with (0002n) peaks highlighted. (**d**) Temperature dependent sheet resistance of 30 nm NC MnTe film from 300 K to 2 K. The hump at 250-300 K is likely related to the AFM transition temperature.

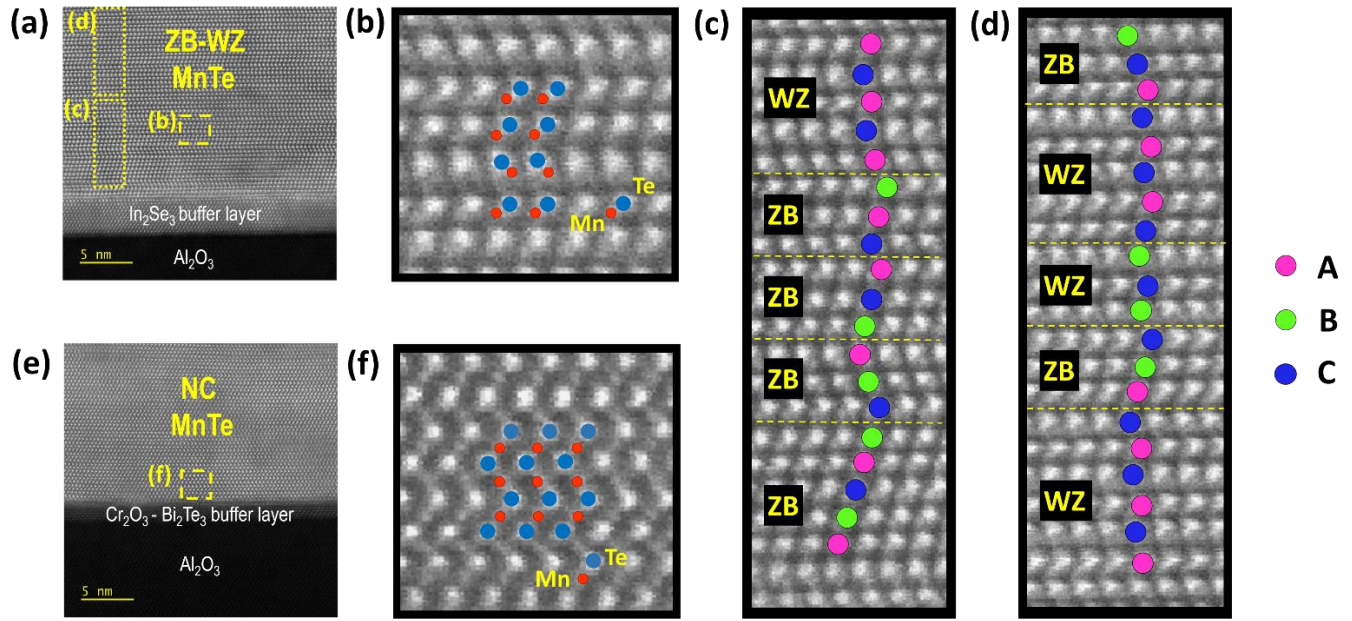

**Figure 5. Cross-sectional HAADF-STEM images.** (a) ZB-WZ MnTe and (e) NC MnTe films with a magnified portion of the films in (b) and (f) respectively, highlighting the difference in arrangement of Mn and Te atoms in both films. (c,d) Transitions in stacking of atoms between ZB and WZ phase during growth of ZB-WZ MnTe as highlighted in (a). A, B and C are the three possible positions (of Te atoms) out of which three are repeated in the ZB structure (ABCABC..) and two are repeated in the WZ structure (ABAB..,BCBC..or CACA). Dashed yellow lines represent stacking faults.